\documentclass[final,5p,times,twocolumn]{elsarticle}

\usepackage{graphics}

\usepackage{amssymb}

\journal{Special Issue: M2S IX}

\begin{document}

\begin{frontmatter}



\title{Comparison of the Hole Concentration determined by Transport Measurement for the Hole-doped Cuprate Superconductors}


\author[label1]{Tatsuya Honma}
\author[label2]{Pei Herng Hor}

\address[label1]{Department of Physics, Asahikawa Medical College, Asahikawa, Hokkaido 078-8510, Japan.}
\address[label2]{Texas Center for Superconductivity and Department of Physics, University of Houston, Houston, TX. 77204-5002, USA.}

\begin{abstract}
We have compared the hole concentration ($P_{pl}$) determined by hole-scale based on the thermoelectric power at RT ($S^{290}$) to the hole concentrations ($P$) determined by two popular hole-scales based on the superconducting critical temperature ($T_c$) and Hall coefficient ($R_H$). While the hole concentrations based on different hole-scales are different, we show that when the $P_{pl}$ is divided by either the effective unit cell volume ($V_{euc}$) which is the unit-cell volume per one CuO$_2$ plane or the optimal hole concentration ($P_{pl}^{opt}$) we can find some correlation between $P_{pl}$ and $P$. That is, the normalized $T_c$ ($T_c$/$T_c$($P_{pl}^{opt}$)) and the Hall number (1/$eR_H$) are well scaled with $P_{pl}$/$P_{pl}^{opt}$ and $P_{pl}$/$V_{euc}$, respectively. We find that the $P_{pl}$-scale can map to and reproduce the other two hole scales if proper dimensionality and normalization are taken into account but not vice versa.
\end{abstract}

\begin{keyword}
Room-temperature thermoelectric power \sep hole-doping concentration \sep Hall number \sep superconducting critical temperature
\end{keyword}

\end{frontmatter}


\section{Introduction}
How to reliably measure the doped hole concentration ($P$) is one of the experimentally important problems for the high temperature superconductor cuprates (HTSC). A working hole scale will allow us to quantitatively compare various physical properties of many HTSC materials to identify intrinsic and universal properties of HTSC.

Based on the fact that La$_{2-x}$Sr$_x$CuO$_4$ (SrD-La214) and Y$_{1-x}$Ca$_x$Ba$_2$Cu$_3$O$_6$ (CaD-Y1236) are two rare HTSC materials in which the $P$ can be uniquely determined from the cation content alone, we have systematically studied the determination and carefully delineate the differences and the importance of the dimensionality of doped hole concentration for the HTSC \cite{hon04,hon06,hon08}. Our analysis indicates that the thermopower at 290 K ($S^{290}$) can be used as a scale for hole-doping concentration per CuO$_2$ plane ($P_{pl}$) which is consistent with both SrD-La214 and CaD-Y1236 \cite{hon04}. Further, we found that the Hall number ($n_H$) determined from the Hall coefficient ($R_H$) is well related with the hole-doping concentration per effective unit cell volume ($P_{pl}$/$V_{euc}$), where the $V_{euc}$ is the unit-cell volume per one CuO$_2$ plane, in the case of single-layer HTSC \cite{hon06}. Furthermore we show that hole concentration of the $P_{pl}$-scale is consistent with that determined by many other macroscopic and microscopic techniques, such as the titration technique, angle-resolved photoemission spectroscopy (ARPES) and near edge x-ray absorption fine structure (NEXAFS) \cite{hon08}.

There are other popular hole-scales proposed. For instance the hole scale based on the dome-shaped $T_c$ behaviour of SrD-La214 was proposed and conveniently used. In this scale, the \textquotedblleft $T_c$-scale\textquotedblright\, $T_c$/$T_c^{max}$ = 1-82.6($P$ -0.16)$^2$ where the $T_c^{max}$ is a maximum in $T_c$ for the HTSC material \cite{pre91}. Also, the inverse Hall number per effective unit cell volume ($n_HV_{euc}$)$^{-1}$ is proposed as the hole-scale \cite{and00}. We call it \textquotedblleft $R_H$-scale\textquotedblright. In this report, we critically compare the differences among these three hole-scales, that is, our proposed $P_{pl}$-scale, $T_c$-scale and $R_H$-scale.

\section{Results and Discussion}

\begin{figure}[t]
\includegraphics[scale=0.5]{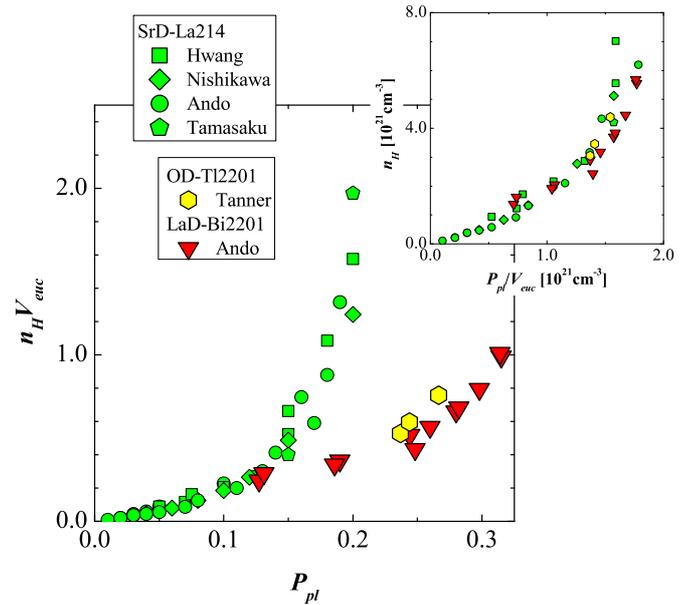}
\caption{\label{fig1} $n_HV_{euc}$ vs. $P_{pl}$ for the typical HTSC materials. The difference between $n_HV_{euc}$ curve for the SrD-La214 and the other materials is the difference between the $R_H$-scale and our $P_{pl}$-scale. The inset shows $n_H$ vs. $P_{pl}$/$V_{euc}$ for the typical HTSC materials.}
\end{figure}

In Figure\ \ref{fig1} we plot the Hall number per effective unit cell $n_HV_{euc}$ as a function of $P_{pl}$ for the typical HTSC. The $n_HV_{euc}$ of the SrD-La214 comes from the refs. \cite{and00,hwa94,nis95,and01,tam94}. The $n_HV_{euc}$ curve for the SrD-La214 exponentially increases with hole-doping. We also plot the data of Tl$_2$Bar$_2$CuO$_{6+\delta}$ (OD-Tl2201) \cite{tan91} and Bi$_2$Sr$_{2-x}$La$_x$CuO$_{6+\delta}$ (LaD-Bi2201) \cite{and00b} as a function of $P_{pl}$. The $n_HV_{euc}$ vs. $P_{pl}$ for the OD-Tl2201 and LaD-Bi2201 deviates upward from the $n_HV_{euc}$ curve of the SrD-La214. In the $R_H$-scale, the $P$ is determined according to the inverse Hall number per effective unit cell ($n_HV_{euc}$)$^{-1}$ curve of SrD-La214 \cite{hon07}. That is, the $n_HV_{euc}$ vs. $P$ or ($n_HV_{euc}$)$^{-1}$ vs. $P$ is assumed to lie universally on a common curve. However, on the $P_{pl}$-scale there is no universal relation in $n_HV_{euc}$ vs. $P$ or ($n_HV_{euc}$)$^{-1}$ vs. $P$. Accordingly, our $P_{pl}$-scale is not consistent with the $R_H$-scale. In stead we found that the Hall number of single-layer HTSC are well scaled by not $P_{pl}$ but $P_{pl}$/$V_{euc}$ \cite{hon06}. The inset shows the Hall number as a function of $P_{pl}$/$V_{euc}$. The Hall number vs. $P_{pl}$/$V_{euc}$ curves for the SrD-La214, OD-Tl2201 and LaD-Bi2201 are found to lie on a common curve as a function of $P_{pl}$/$V_{euc}$ \cite{hon06}. This suggests that our $P_{pl}$/$V_{euc}$ can be related with not $n_HV_{euc}$ but $n_H$. The Hall number or $n_H$ is conventional 3D hole concentration. It is consistent with $P_{pl}$/$V_{euc}$. On the other hand, the $P_{pl}$ or $n_HV_{euc}$ is essentially 2D hole concentration, while the $P_{pl}$/$V_{euc}$ or $n_H$ is 3D hole concentration. If the material is homogeneous and isotropic then the $P_{pl}$ and $P_{pl}$/$V_{euc}$ or $n_H$/$V_{euc}$ and $n_H$ cannot be distinguished. Accordingly, the difference in the physical meaning between $n_HV_{euc}$ vs. $P_{pl}$ and $n_H$ vs. $P_{pl}$/$V_{euc}$ is the difference in the characteristic dimensionality of the physical properties measured by thermoelectric power and Hall coefficient. So the our $P_{pl}$/$V_{euc}$ is the 3D version of the $P_{pl}$-scale.

\begin{figure}[b]
\includegraphics[scale=0.5]{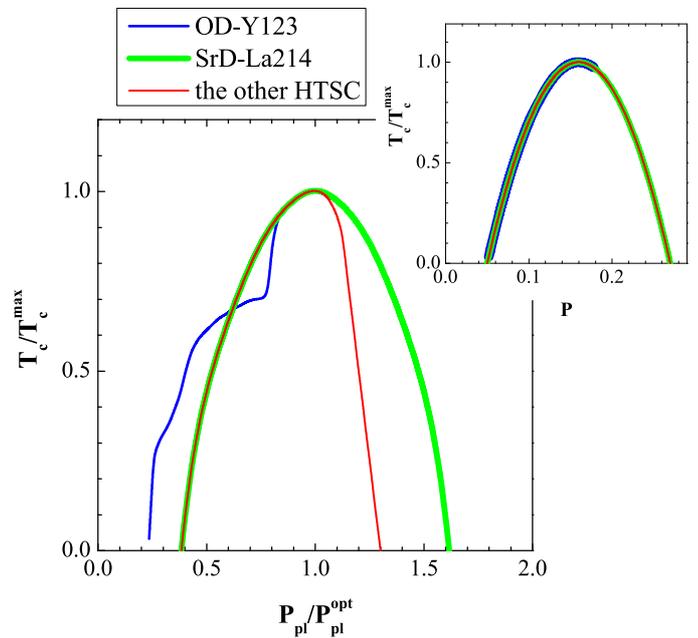}
\caption{\label{fig2} $T_c$/$T_c^{max}$ vs. $P_{pl}$/$P_{pl}^{opt}$ for the HTSC. The green curve is for SrD-La214. The blue curve is for the OD-Y123. The red curve is for the other HTSC. The inset shows the $T_c$/$T_c^{max}$ vs. $P$ determined by $T_c$-scale.}
\end{figure}

Figure\ \ref{fig2} summaries the $T_c$/$T_c^{max}$ as a function of the $P_{pl}$ divided by the optimal hole concentration ($P_{pl}^{opt}$) where $T_c$ = $T_c^{max}$ \cite{hon08}. The $T_c$/$T_c^{max}$ for the SrD-La214 shows the well-known dome-shaped $T_c$-curve \cite{pre91}. The YBa$_2$Cu$_3$O$_{6+\delta}$ (OD-Y123) shows a double-plateau corresponding to the 60 K and 90 K phases \cite{hon07}. The other HTSC, which is almost all major HTSC except of SrD-La214 and OD-y123, follows an asymmetric half-bell-shaped $T_c$-curve \cite{hon08}. When we plot all the data on the $T_c$-scale, all structure appeared on the $P_{pl}$-scale collapse into the dome-shaped $T_c$-curve as shown in the inset. Except of OD-Y123, the $T_c$/$T_c^{max}$ curve in the underdoped side shows the similar behavior on both scales. Therefore , in the underdoped regime, the $T_c$-scale is same as the $P_{pl}$-scale which is normalized to the optimal doping concentration. Accordingly, the $T_c$-scale in the underdoped side is nearly identical to $P_{pl}$/$P_{pl}^{opt}$ based on the proposed $P_{pl}$-scale, but the $T_c$-scale in the overdoped side is not consistent with the $P_{pl}$-scale. Many HTSC materials were studied and compared in the underdoped side. This is propably why the $T_c$-scale seems to work plausibly in many doping dependence studies where only the trend, not the absolute doping concentration, of the doping dependence was the primary concern.

In summary, we have compared our proposed hole-scale based on the $S^{290}$ with two popular hole-scales. The hole concentrations determined by three hole-scales are not consistent among each other. However the normalized $T_c$ ($T_c$/$T_c$($P_{pl}^{opt}$)) and the Hall number ($n_H$) are well scaled with $P_{pl}$/$P_{pl}^{opt}$ and $P_{pl}$/$V_{euc}$, respectively. Therefore, the $P_{pl}$-scale can map to and reproduce the other two hole scales if proper dimensionality and normalization are taken into account but not vice versa.

\end{document}